\documentclass[a4paper,11pt]{article}
\usepackage{jinstpub} 

\title{\boldmath Track reconstruction for the COMET Phase-II experiment with ACTS}

\author[1]{Amaia Razquin\note{Current affiliation: Institute of Physics, University of Graz, 8010 Graz, Austria.}}
\author[2]{and MyeongJae Lee\note{Corresponding author.}}
\affiliation{Department of Physics, Sungkyunkwan University,
Suwon, 16419, Republic of Korea}

\emailAdd{myeongjaelee@skku.edu}

\abstract{

An implementation of A Common Tracking Software (ACTS) toolkit for signal electron reconstruction for the COMET 
muon 
to electron 
conversion experiment is discussed. 
The COMET experiment in J-PARC, Japan, will search for neutrinoless conversion of muons into electrons in the field of an aluminium nucleus, a lepton flavour violating process,
aiming target sensitivity of $10^{-17}$. 
To achieve its scientific goals, 
the experiment
requires a reconstructed momentum resolution of lower than 
150 keV/c.
For the first time by applying ACTS to signal events in the 100 MeV energy range with multiple-turn trajectories in the presence of background events, it is found that
the reconstruction efficiency is around 14\% with no fake reconstructed events. The implementation details, performance, and issues of ACTS in the context of COMET are presented.}


\begin{document}
\maketitle
\flushbottom

\section{Introduction}
\label{sec:intro}
The COMET experiment~\cite{COMET:2018auw} will search for neutrinoless muon to electron conversion in the field of a nucleus 
at J-PARC, Japan. Muon to electron conversion, namely $\mu^-+N(A,Z)\rightarrow e^-+N(A,Z)$, is one of the most important charged lepton flavour violating (CLFV) processes \cite{Bernstein:2013hba}. This transition is highly suppressed in the Standard Model (SM) with a branching ratio of $\mathcal{O}(10^{-54})$, but many extensions of the SM predict transition rates that may reach observable levels \cite{Kuno:1999jp}. The current limits of muon to electron conversion were set by the SINDRUM-II experiment in 2006 
with Gold target
at $7\times10^{-13}$ \cite{SINDRUMII:2006dvw}. COMET aims to improve this by a factor of 
$100 \sim 10,000$
in Phase-I and Phase-II, respectively. 

\begin{figure}[htbp]
\centering
\includegraphics[width=.8\textwidth]{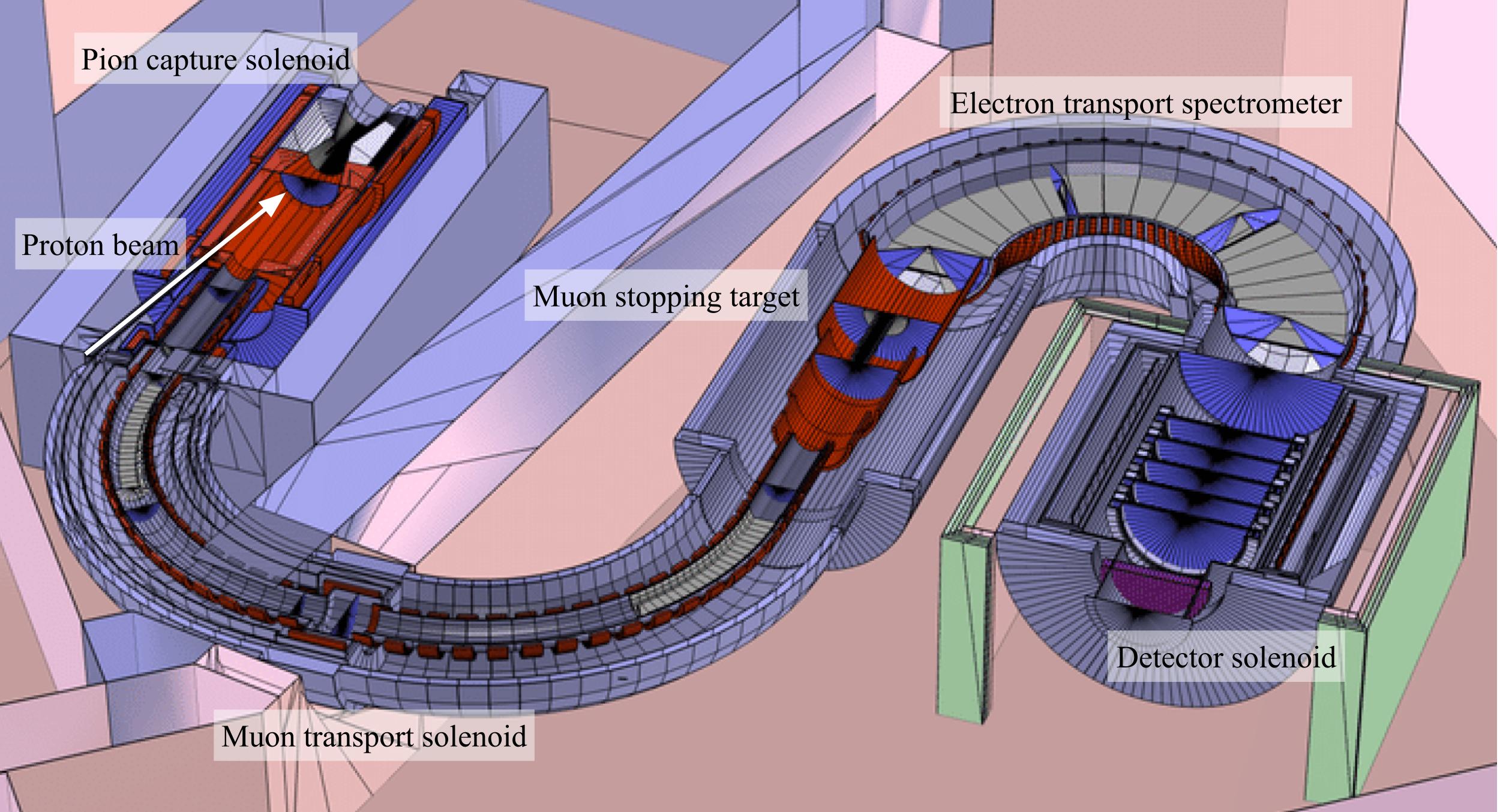}
\caption{Layout of the COMET Phase-II experiment.\label{fig:layout}
The proton beam from J-PARC enters the Pion-capture solenoid, shown in the left side of the figure. The pions and its decay muons are transported through the Muon-transport solenoid, before arriving the Muon stopping target, shown in the middle of the figure. The signal or background electrons pass the Electron transport solenoid, and are measured by the straw tracker and calorimeter detector located inside the Detector solenoid, shown in the right side of the figure. 
}
\end{figure}

A schematic layout of Phase-II~\cite{Benjamin} 
is shown in Fig.~\ref{fig:layout}. The experiment begins with a 56 kW proton beam from the J-PARC main ring hitting a tungsten target to produce pions. The pions captured by the transport solenoid decay into muons and reach the muon stopping target region. A fraction of the muons is stopped in the target and becomes atomically bound to the Coulomb potential of the nucleus. The muons subsequently cascade down to the $1s$ orbital. Three processes can occur in the muonic atom: (1) the muon is captured by the nucleus, (2) the muon decays in orbit (DIO) without interacting with the nucleus, or (3) the muon 
conversion
occurs
via a CLFV process. COMET Phase-II will have a muon stopping target made of aluminium. The lifetime of an aluminium muonic atom is 864~ns, and the ratios for muon nuclear capture and DIO are 61\% and 39\%, respectively. In the case (3), neutrinoless coherent muon to electron conversion can occur, where the outgoing electron is a mono-energetic electron of 104.97~MeV 
(in the case of an Aluminium target). The energy of the electron is given by the muon mass minus the $1s$ orbital binding energy and the recoil energy of the nucleus. The converted electrons are transported through the electron spectrometer until they reach the detector solenoid. In Phase-II, the detector will be composed of a straw tracker and an electromagnetic calorimeter.  
The construction of COMET Phase-II experiment starts after the Phase-I completion.  
The COMET Phase-I experiment, which starts data taking in FY2025, employs the same beamline for proton and pion production as Phase-II, but the length of the muon transport solenoid is half of the length of the solenoid in Phase-II. This is followed by a Cylindrical drift chamber detector for measuring the muon conversion.

The particles arising from the proton beam and muon DIO electrons compose the main backgrounds of the COMET Phase-II experiment. The beam-related backgrounds are suppressed by utilising two C-shaped transport solenoids and a proton beam with an extinction factor of $10^{-11}$ \cite{Noguchi:2022ioy}. The DIO electrons are produced alongside two neutrinos (
$\bar{\nu}_e$
and $\nu_\mu$). The electron energy spectrum follows a Michel decay spectrum, peaking at around half the muon mass (52.83~MeV).
There is a high-energy tail extending up to the end-point energy, which is close to 104.97~MeV, attributed to the nuclear recoil.
Therefore, the detector requires a momentum resolution of better than 150~keV/c to distinguish the signal electron from DIO backgrounds.

The track reconstruction algorithms in COMET Phase-II have to be able to effectively reconstruct the multi-turn trajectories of signal electrons in the straw tracker. A Common Tracking Software (ACTS) \cite{Ai:2021ghi} is a track reconstruction toolkit based on the tracking algorithms in the ATLAS experiment. ACTS aims to be an experiment-independent and framework-independent toolkit for high energy physics experiments. It has been implemented in experiments such as ATLAS \cite{ATLAS:2023iat}, sPHENIX \cite{Osborn:2021zlr}, FASER \cite{FASER:2023zcr}, CEPC \cite{Chen:2023abc} and STCF \cite{Ai:2023ukc}, among others. Thus, the COMET collaboration considers it as a candidate for reconstruction in the Phase-II experiment. 
The energy range of these experiments is in the GeV scale, and most detectors are composed of a central barrel-like detector that particles traverse radially outwards, which is not the case for COMET Phase-II experiment where the particles of 100~MeV range traverse the solenoidal field with multiple-turn trajectories.   
Therefore, 
the implementation of ACTS in COMET Phase-II represents a novel application of ACTS in the 100~MeV energy range for multiple-turn tracking. 

In this paper, the implementation of ACTS into the straw tracker of COMET Phase-II and its performance will be presented. Additionally, the applicability and future needs of ACTS in the context of COMET will be discussed. The article is organised as follows. In section \ref{sec:detector} the detector of Phase-II is presented. The implementation of ACTS is described in section \ref{sec:reconstruction}. In section \ref{sec:performance} the track reconstruction performance in the presence of background events is presented. A conclusion alongside the issues encountered are described in section \ref{sec:conclusion}.

\section{The COMET Phase-II detector}
\label{sec:detector}

The detector in COMET Phase-II is a combination of a straw tracker and an electromagnetic calorimeter (ECAL). The straw tracker aims to measure the particle momentum with high resolution in order to identify the incoming particles. In particular, it must have high momentum resolution for signal electrons at 105~MeV/c. The straw tracker is positioned inside a 1~T solenoid field and it is composed of super-layers, or stations, evenly distributed through the axis of the magnetic field. Each station consists of two double-arrays, or manifolds, 
one double-array with the straws vertically positioned and the other one horizontally positioned to measure the $x$ and $y$ coordinates, respectively.
Within each manifold, the straws are positioned in two layers staggered by half a straw diameter in order to solve left-right ambiguities. 

The straws are made of a compound of mylar and aluminium, with a thickness of 12.5~$\mu$m and a radius of 4.9~mm. The length of the straws in every layer goes from 79 cm to 139 cm in groups of varying lengths. The anode wires inside the straw tubes are made of gold-coated tungsten and a gas mixture of 50\%-Ar and 50\%-C$_2$H$_6$
is provided. The gap between the stations is kept at a $~0.1$~Pa vacuum. 

The ECAL serves three main purposes: measure the energy of electrons with good resolution, add redundancy to the momentum measurement, and provide the E/p ratio for electron identification. It is positioned downstream of the detector solenoid, after the last of the straw tube stations. The unit of the ECAL is a $2\times2$ segmented scintillating crystal matrix. The crystals are made of LYSO. The whole calorimeter will consist of 500+ modules in a cluster position. The ECAL will be used to achieve an energy resolution better than 5\% at 105~MeV and a cluster position resolution of 1~cm.

The straw tracker spatial resolution is 0.15~mm and the timing resolution is better than 2~ns \cite{Nishiguchi:2020yly}. The momentum resolution required for Phase-II is 150~keV/c. To achieve that, tracking algorithms that accurately reconstruct the helical trajectories of electrons through the straw stations are required.

\section{Straw Tracker reconstruction using ACTS}
\label{sec:reconstruction}

\subsection{ACTS program}
\begin{figure}[b!]
    \centering
    \includegraphics[width=0.8\textwidth]{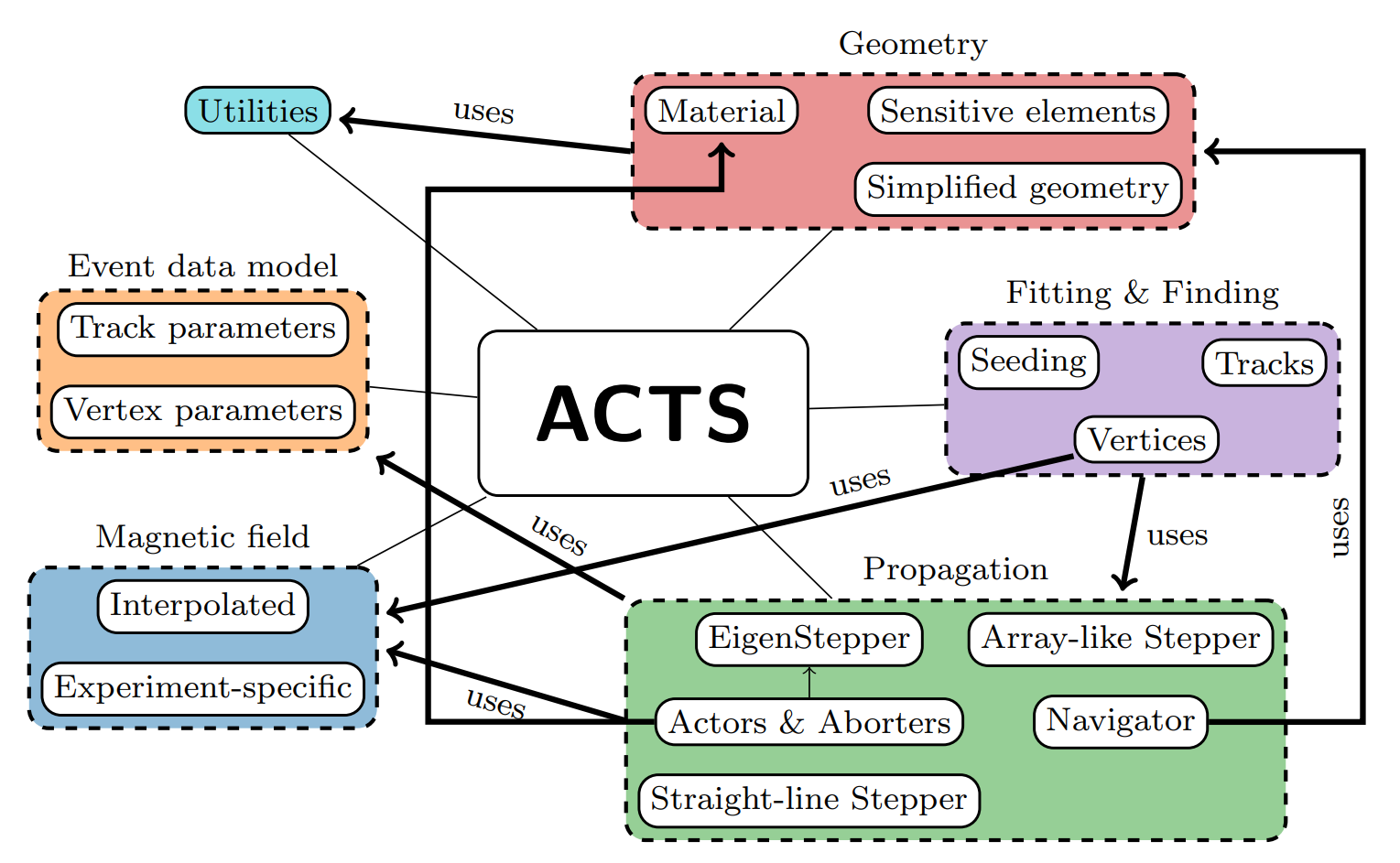}
    \caption{ Overview of selected components of ACTS and their cross-dependencies, from Ref. \cite{Ai:2021ghi}.
The components (such as propagation and geometry) are categorised into modules, and shown as coloured boxes. }
    \label{fig:acts}
\end{figure}

ACTS is a high energy physics software, written in modern C++17 and with minimal dependencies.
The main goals of ACTS are: 
(a) to preserve
and update the well-tested code from the LHC experiments and enable it for preparation
to the High Luminosity era of the LHC; 
(b) to provide a development test bed
for algorithmic research and portability to accelerated hardware; and
(c) to provide a mature track reconstruction toolkit applicable to any experiment.
The components for track reconstruction within ACTS are categorised in modules, as shown in Fig.~\ref{fig:acts}. 
The essential modules are the event data model, propagation, geometry,
magnetic field, and fitting and finding modules. A detailed introduction to these modules can be found in Ref.~\cite{Ai:2021ghi}.

\subsection{Geometry and magnetic field}
\begin{figure}[b!]
    \centering
    \begin{minipage}[b]{.48\linewidth}
        \centering
        \includegraphics[scale=0.12]{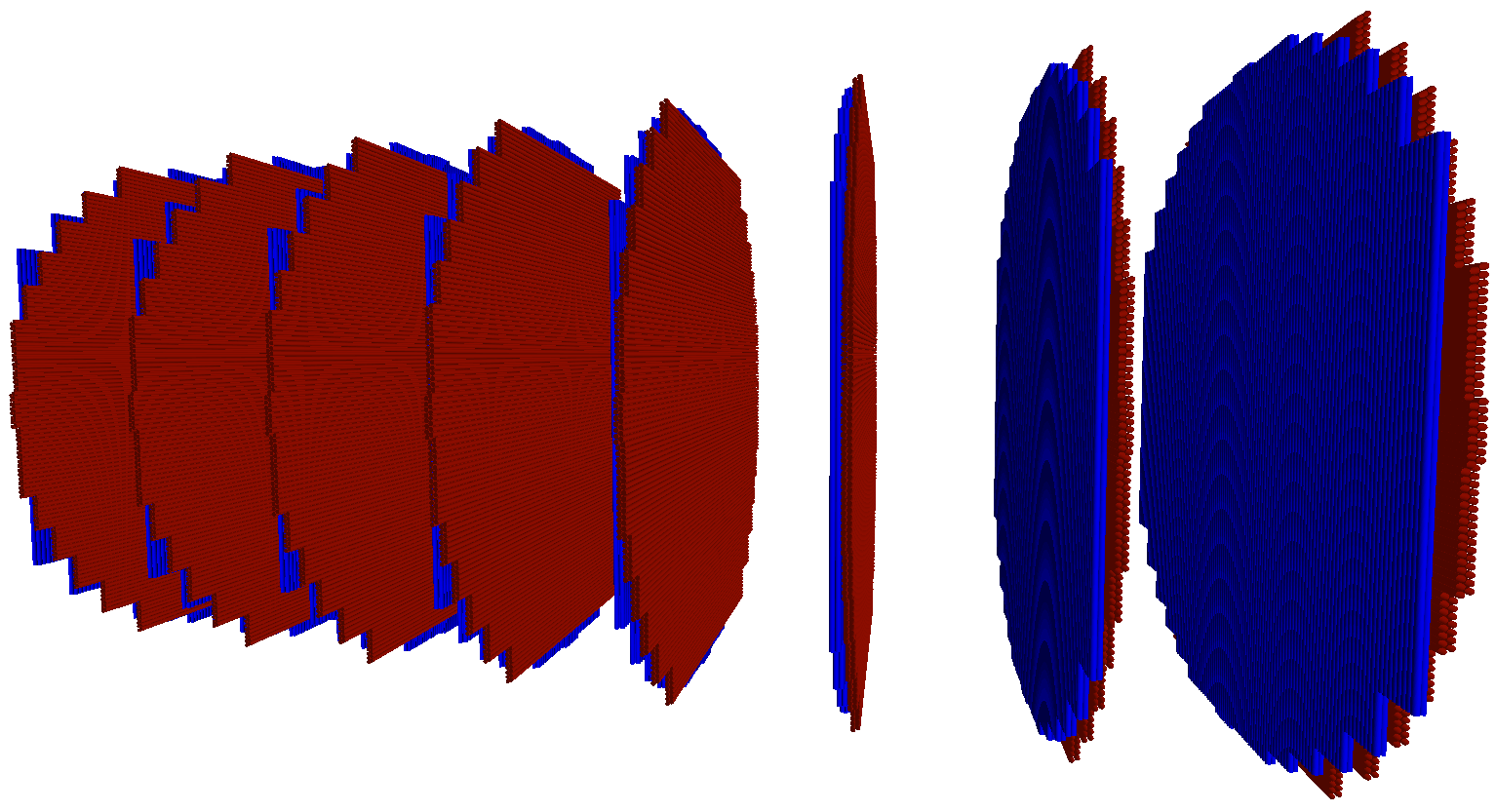}
    \end{minipage}
    \begin{minipage}[b]{.48\linewidth}
         \centering
        \includegraphics[scale=0.122]{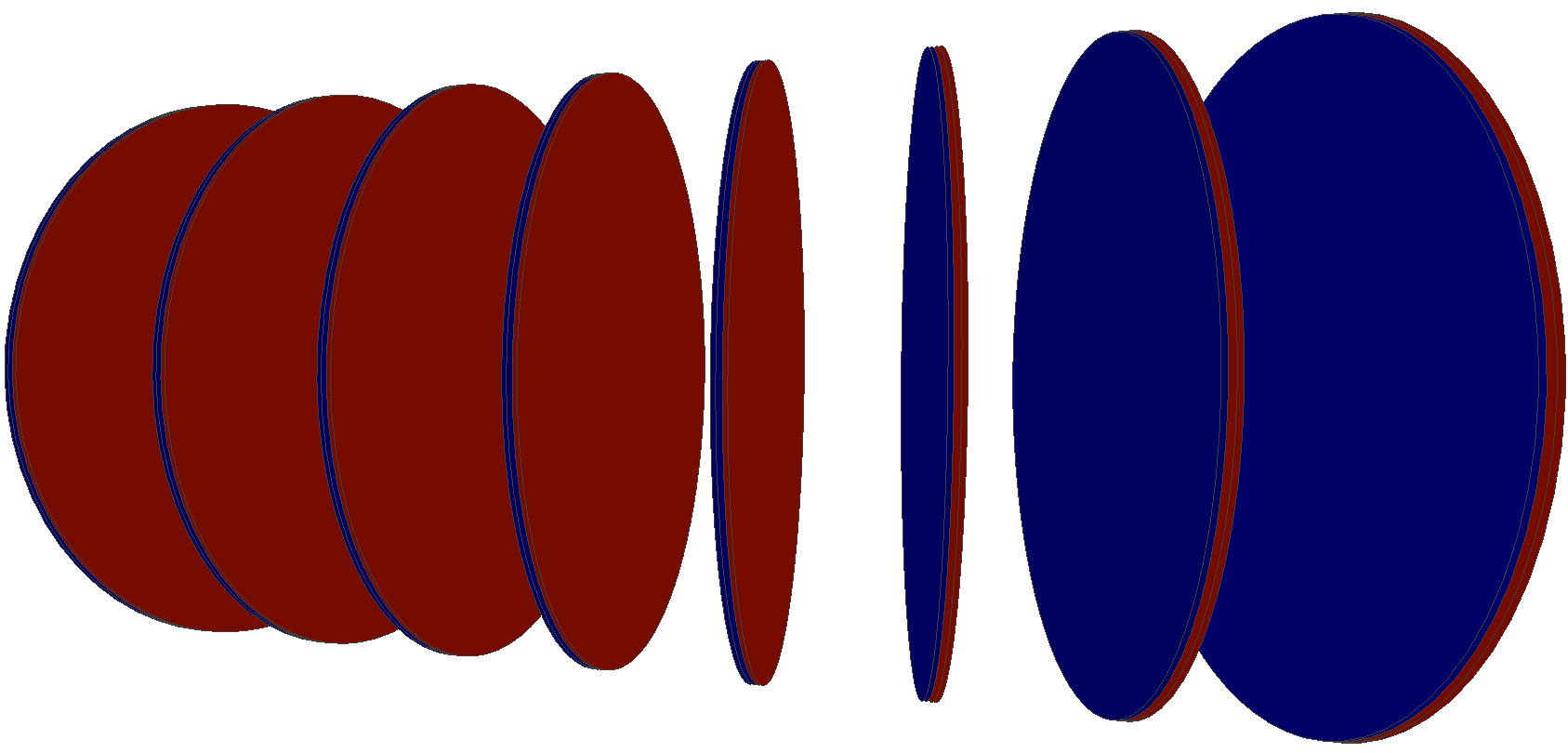}
    \end{minipage}\par\medskip
    \caption{The two geometries of the COMET Phase-II experiment implemented in ACTS: the detailed straw geometry (left), and the disc simplified geometry (right).}
    \label{fig:geometry}
\end{figure}

The geometry description in ACTS is a simplified version of the real geometry. All the elements in the geometry are based on the \texttt{Acts::Surface} class. The detector element surfaces are gathered in \texttt{Acts::Layer}s, and the layers are embedded within \texttt{Acts::Volume}s. ACTS contains a ROOT \cite{Brun:1997pa} \texttt{TGeometry} plugin that can be used to convert the detailed Geant4 \cite{Allison:2016lfl} COMET Phase-II geometry into ACTS geometry. However, the \texttt{Acts::Layer}s overlap when introducing staggered straw layers. Thus, at the moment the Phase-II geometry is manually coded into ACTS. Two geometries were built to resemble the Phase-II geometry: (a) a simplified ``disc" geometry, and (b) a realistic ``straw" geometry (see Fig.~\ref{fig:geometry}). 

The disc geometry (a) was constructed by substituting the straw layers with disc layers. Thus, every station contained four disc surfaces with $r=695$~mm. The material of the discs is the same Mylar compound as the straw tubes, and the thickness is double the thickness of the tube walls. The material properties were extracted with the \texttt{TGeometry} plugin. 

The straw geometry (b) was built by modelling every straw following the real position and orientation. All the straws in a layer were embedded in an \texttt{Acts:Layer}, with the layer thickness being slightly smaller than the straw thickness to avoid layer overlapping. The material was mapped accurately using the \texttt{TGeometry} plugin. 

In ACTS, the measurements need to be associated with detector surfaces. A ROOT-based reader was implemented to convert the simulated Geant4 hits to ACTS measurements. For the disc geometry, the hits were mapped into the surface as two-dimensional measurements, ($r$,$\phi$), representing the radial distance from the centre of the disc and the azimuthal angle in the XY plane, respectively. In the realistic straw geometry, the measurements are one dimensional, $l_D$, and they represent the drift distance of the ionized electrons to the anode wire within the straw tube. 

To perform charged particle reconstruction, the magnetic field within the detector is required. The discrete values of the solenoidal magnetic field in the straw tracker are introduced via a ROOT file into the ACTS interpolated magnetic field. ACTS performs a cell-wise interpolation to get the values of the field at each propagation step. 

\subsection{Reconstruction workflow}
\begin{figure}[htbp]
\centering
\includegraphics[width=.7\textwidth]{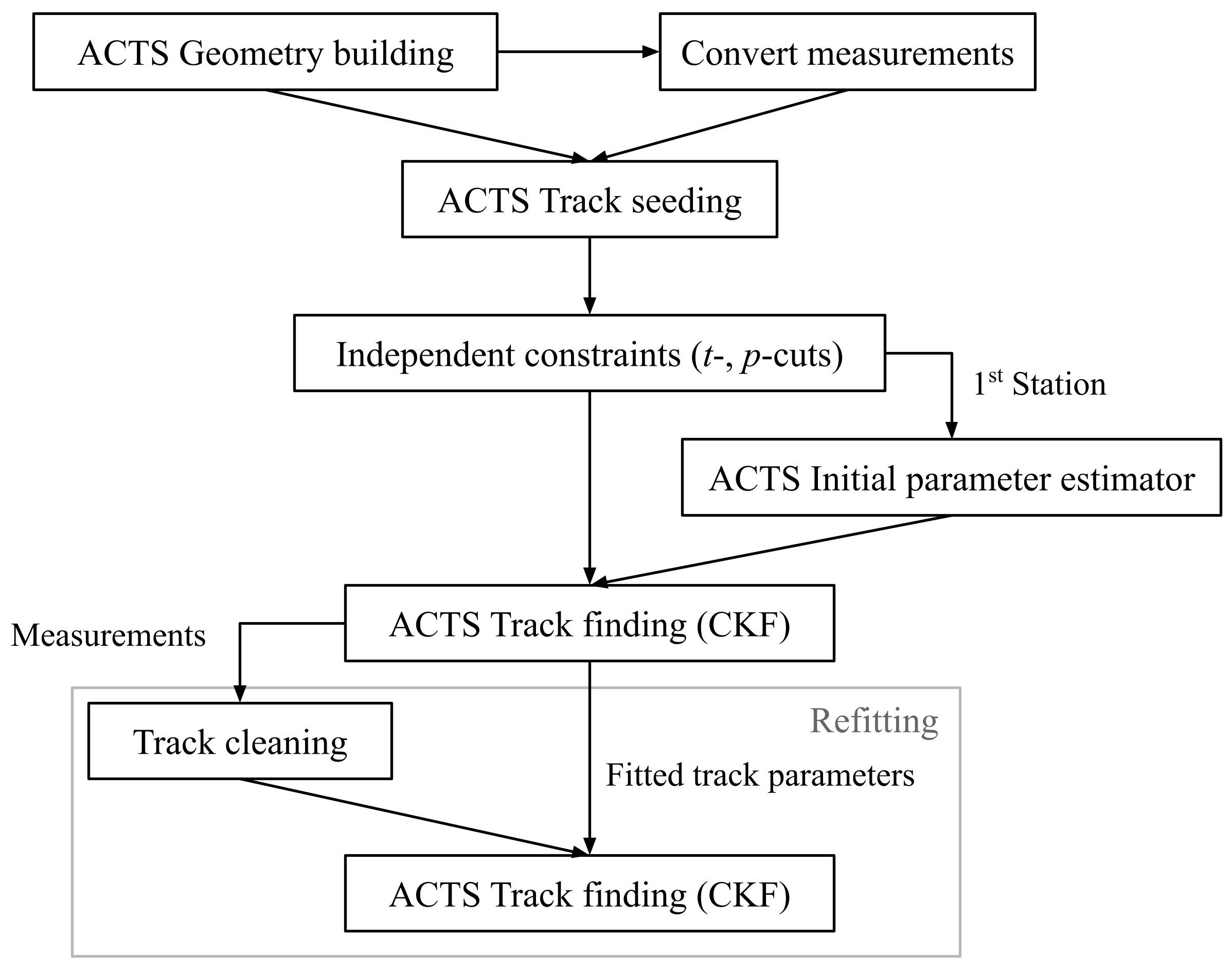}
\caption{Workflow of track reconstruction for the straw tracker in COMET Phase-II using ACTS.\label{fig:workflow}}
\end{figure}

The track reconstruction workflow can be found in Fig.~\ref{fig:workflow}. The reconstruction occurs entirely within the ACTS framework and uses several ACTS tools alongside dedicated COMET Phase-II algorithms. After creating the geometry, adding the magnetic field, and linking the measurements, the reconstruction begins by seed finding in each station. Seed finding, within ACTS, is the process of finding three measurements (a triplet) believed to belong to the same trajectory. The seed finder algorithm within ACTS is used for COMET Phase-II. Seeds are formed exclusively with measurements within the same station. In addition, the triplets undergo two independent constraints in order to remove background and fake seeds. The first constraint, $t$-cut, considers the detection time of the measurements within the triplet. If a hit is detected too early with respect to the previous hit, the triplet is deemed fake. For this calculation, the maximum drift time ($t_D^{\text{max}}=100$~ns) and the time resolution of the straw triggers ($t_{\text{resolution}}=2$~ns) are considered: 
\begin{equation} \label{eq:tcut}
    t(z_1)-t(z_2)>t_{\text{resolution}}+t_D^{\text{max}},
\end{equation}
where $t(z_1)$ and t($z_2)$ are the detection times of hits at position $z_1$ and at $z_2$ with $z_1<z_2$, respectively. If eq.~\ref{eq:tcut} is true, the triplet is discarded. 

The second constraint, $p$-cut, uses the ACTS initial parameter estimator from seeds to approximate the momentum of the triplet. The triplets with a momentum estimation that deviates considerably from the expected estimation for signal electrons are removed. The two independent cuts are applied in order to remove background measurements. Thus, only the measurements belonging to the remaining triplets are kept in the next steps of track reconstruction. To avoid removing signal electron measurements, the $p$-cut is applied considering a previous confidence scan. A perfect trajectory in a station will leave one hit in each straw layer, a total of four hits per station. Consequently, since the triplets are formed with measurements in the same station, four distinct triplets can be found per particle. Thus, if four measurements are grouped in distinct groups of three, each group (triplet) will have two shared measurements with any other group.
The $p$-cut applies less stringent constraints to triplets that share two measurements with four other triplets, as they are deemed to have more probability of belonging to a real trajectory.

The triplets found in the first station and their estimated initial parameters are used to initialise track finding. Track finding is performed by using the Combinatorial Kalman Filter (CKF) included in ACTS. The tracks are found among the measurements that were included in triplets in the track seeding stage. The CKF simultaneously fits the measurements that it has found. At the end of the finding and fitting stage, only the final trajectories with more than 15 measurements are considered. The measurements with a large $\chi^2$ distance in these trajectories are removed (track cleaning) and the remaining ones are fed into the track finder again. In the final track finding run, the fitted parameters are used as initial parameters.

\section{Performance}
\label{sec:performance}
The signal electron and background measurements used in this study were produced in ICEDUST, the main software suite used in COMET based on Geant4. The background events were simulated starting from $2.7\times10^{7}$ protons hitting the pion production target. The background hits in the detector were filtered considering the arrival time and energy deposition following a realistic scenario. The signal events were simulated starting from a distribution of stopped muons in the target and assuming they are created in isotropic directions. In the following sections, the signal electrons are reconstructed individually, with only one signal alongside the background per run.

\subsection{Track seeding}
Track seeding in ACTS is performed by finding triplets of measurements thought to belong to the same trajectory. The standard seed finder in ACTS applies several constraints iteratively to find acceptable triplets between the measurements. These constraints, or ``cuts", have to be tuned for each experiment. For COMET Phase-II, a preliminary analysis of the constraints and their values for signal electrons was performed. The results of the analysis were used to set the ranges for each constraint in an optimisation run. The values were optimised by using the Optuna hyperparameter optimisation software \cite{optuna_2019}. The following objective function was maximised with the automated search in Optuna: 
\begin{equation}
    \text{Score function}=\text{Efficiency}-\frac{\text{Run time}}{W_t},
\end{equation}
where the efficiency is the ratio between the events with at least one seed found and the total amount of events. $W_t$ is a weight factor for the running time. The score is calculated from the performance of ACTS with a set of constraints within the given range and introduced into Optuna alongside the constraints employed. Optuna then proposes new values for the constraints and a new run of ACTS is performed. By doing this process iteratively Optuna finds improved combinations of constraints for maximising the efficiency of ACTS. The straw stations in the Phase-II detector are far apart, in consequence, the seeding algorithm is implemented in each station separately. In this study the same constraints were used for all the stations, thus the best constraints were those that improved the efficiency of seeding in all the stations simultaneously. In a future study different sets of constraints should be considered for each station, since the particles' energy decreases through the detector. After 500 trials, the seed finder optimisation reached an efficiency of 92\% in all stations. 

After tuning the constraints within the seed finder, it was implemented alongside the $t$-cut and $p$-cut above mentioned. Separate runs with 16,000 signal events with realistic background events were performed for (1) the seed finder alone, (2) the finder and one additional cut, and (3) the finder with the two independent constraints. The results for all stations together are shown in Table~\ref{tab:seeding}. A seed is considered fake if the three measurements do not belong to the same trajectory. If they belong to the same trajectory, but the particle they belong to is not a signal electron, the seed is considered a true background seed. The table shows that the additional cuts, especially the $p$-cut, remove up to 60\% of fake and background seeds in comparison to the seed finder algorithm alone. However, the momentum cut also results in a drop in the number of signal seeds: the number goes down from 82\% using ACTS alone, to 74\% using both of the additional constraints.  

\begin{table}[htbp]
\centering
\caption{Comparison of seed finding efficiency,  true background rate, and fake rate, in all the stations together between the ACTS seed finder alone and the seed finder alongside the independent cuts applied for COMET Phase-II. The efficiency, background and fake rates and removal of fake seeds are defined in the text. \label{tab:seeding}}
\smallskip
\begin{tabular}{lcccc}
\hline\hline
& ACTS & ACTS + $t$-cut& ACTS + $p$-cut & ACTS + $t$-cut + $p$-cut\\
\hline
Efficiency & 0.82 & 0.82 & 0.74 & 0.74\\
True background rate & 0.56 & 0.59 & 0.54 & 0.55\\
Fake rate & 0.35 & 0.32 & 0.33 & 0.32\\
Removal of fake seeds &  & 14\% & 60\% & 62\% \\
\hline\hline
\end{tabular}
\end{table}

\begin{figure}[htbp]
\centering
\includegraphics[trim=100 0 180 50, clip, width=.8\textwidth]{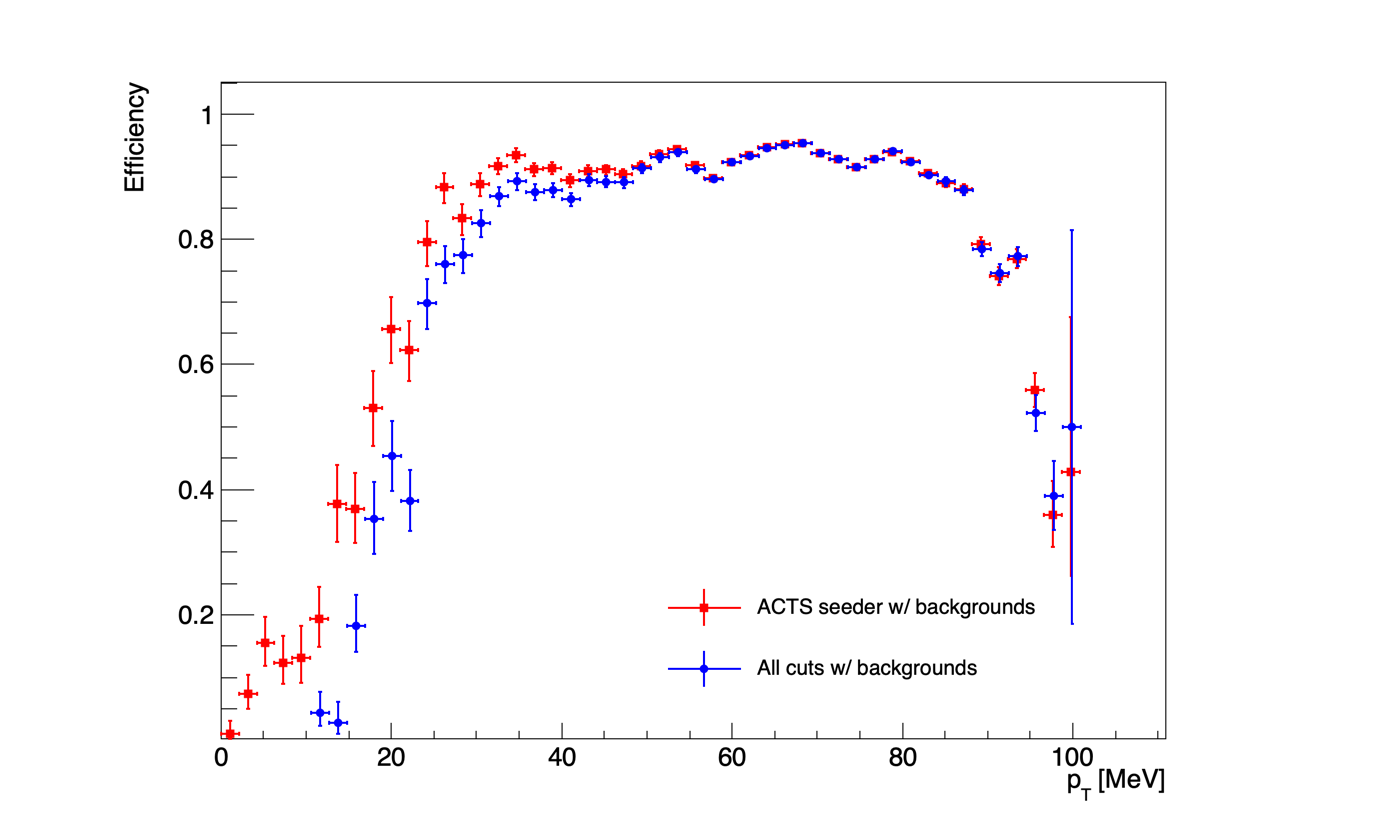}
\caption{Efficiency of track seeding with respect to the transverse momentum. The values for seeding 
with backgrounds and additional base cuts 
(red square marker),
and with backgrounds and all applicable cuts 
(blue circle marker)
are shown.}
\label{fig:seeding_station0}
\end{figure}

Seed finding in the first station is critical for the proper initialisation of track finding. The efficiency of the ACTS seeder, the $t$-cut and the $p$-cut together is 90\% in the first station, with an average of 2.9 seeds per signal track. Without the independent constraints, the efficiency is 90.7\%, but there are 11 times more non-signal initial seeds. The efficiency can be seen in Fig.~\ref{fig:seeding_station0}. The fluctuations in the efficiency are given by the greater presence of background events in those transverse momentum regions. A sharp drop can be seen after $p_T=90$~MeV/c, this occurs because the radial distance between measurements is limited within the seeder. This radial distance is limited in order to avoid measurements from different particles to be combined in the same triplet. In consequence, since radial distance between measurements of the same particle increases with $p_T$, triplets from particles with high transverse momentum are not formed. The value for the maximum radial distance is chosen based on the aforementioned optimisation runs as a trade off between efficiency and background removal. Thus, the independent constraints successfully remove backgrounds without excessively erasing signal measurements.

\subsection{Initial track parameter estimation}
\begin{figure}[htbp]
\centering
\includegraphics[trim=0 0 100 70, clip, width=.8\textwidth]{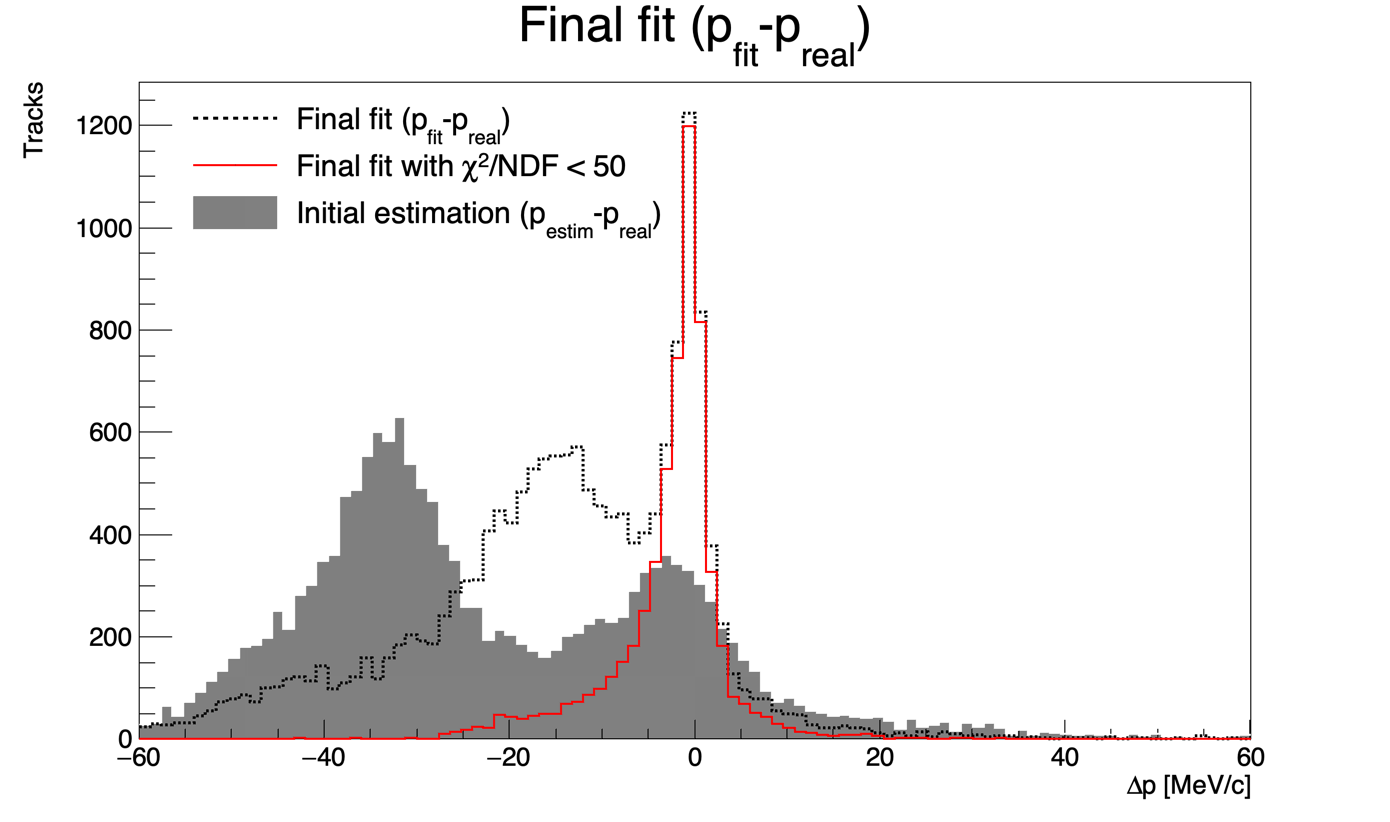}
\caption{
Estimated momentum (grey filled histogram) and fitted momentum (solid red and dotted black histogram) with respect to the real momentum. The estimated momentum is often less than the real momentum which causes a decrease in the fit resolution. A $\chi^2$ cut can successfully remove the results with worse deviation. The portion of fitted momentum with $\chi^2/NDF<50$ is shown in red solid line histogram.\label{fig:init_mom}
}
\end{figure}

The seeds in the first station are required to estimate the initial parameters of the reconstruction. ACTS contains an initial parameter estimator from seeds based on the ATLAS initial estimator. It performs a conformal transformation to extract the trajectory parameters. This estimator is designed with barrel detectors in mind, which is not the case for the straw tracker in COMET Phase-II experiment. Nevertheless, it was implemented alongside the seeder. The estimated initial momentum deviation with respect to the real momentum is shown in Fig.~\ref{fig:init_mom}. It can be seen that the estimator excessively miscalculates the momentum of the trajectory, reaching a deviation of 70\% in some cases. 

An improper setting of the initial parameter estimator can cause a decrease in the accuracy of the track fitting. To address this issue, circle fitting algorithms were considered. The Taubin fit \cite{Taubin1991Estimation}, the Hyper fit \cite{al2009error} and the Levenberg-Marquardt fit \cite{Levenberg} 
were implemented, together and separately. None of them yielded better results than the ACTS estimator, because the arc that the measurements in a triplet subtend is very small. Additionally, the estimations by the ACTS algorithm were used as initial values in a circle inexact line search minimization, but this also did not show improvements with respect to the original values. In consequence, the initial parameter estimator within ACTS is better than circle fit estimators, but it is often insufficient for the energy range and resolution requirements of COMET Phase-II. 

\subsection{Track finding and track fitting}
Track finding is the process of grouping all of the measurements that belong to the same trajectory. Track fitting is the process of fitting those measurements to extract the parameters of the original trajectory. The CKF combines the two procedures by iteratively fitting the measurements found and discarding the ones that deviate from the expected trajectory. The fitting is done using a Kalman Fitter, and the measurements are discarded based on the $\chi^2$ distance between the hit and the prediction of the fitter.

\begin{figure}[t!]
\centering
\includegraphics[trim=0 0 100 50, clip, width=.8\textwidth]{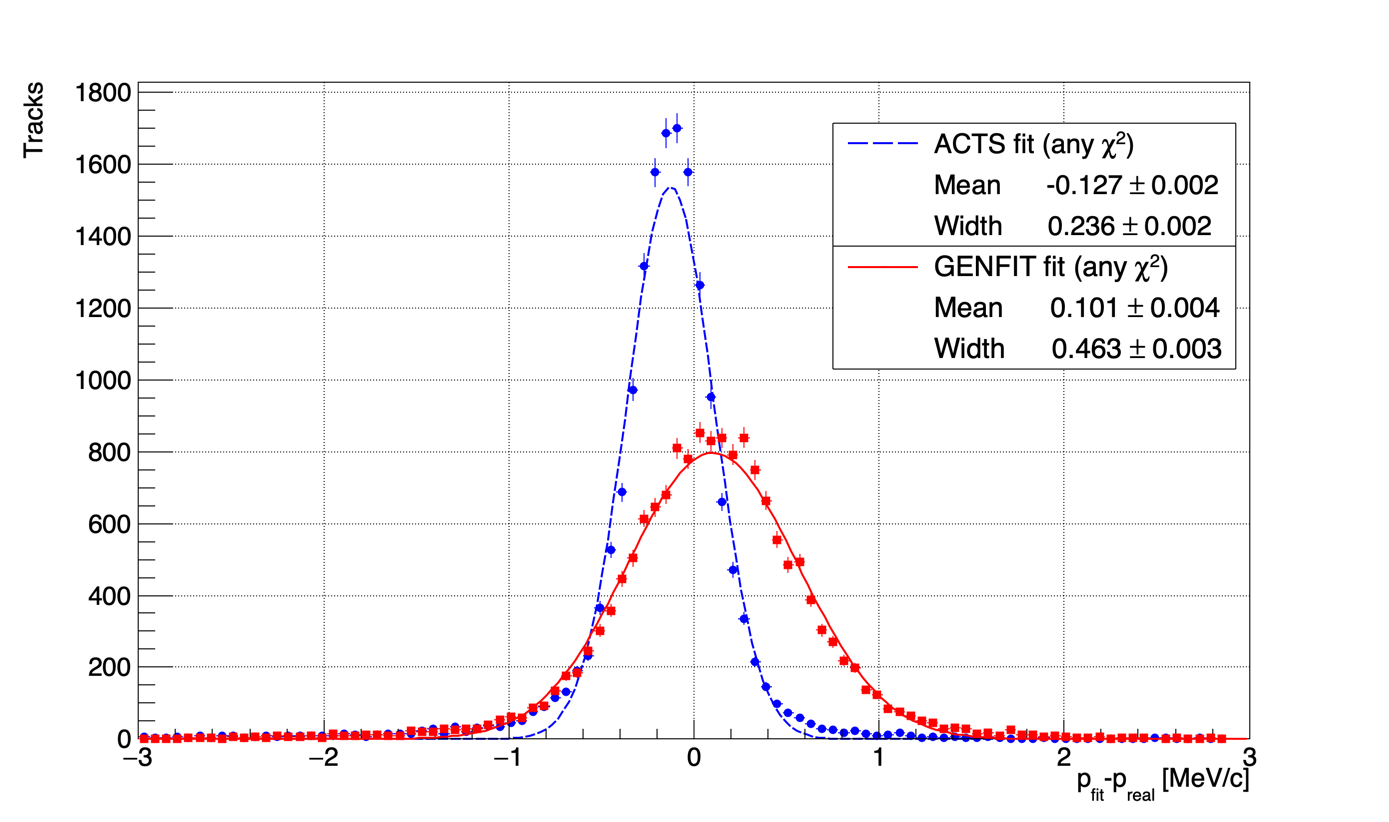}
\caption{
The reconstructed momentum difference from its real value are shown. The blue circle points are the momentum data from ACTS reconstruction and its Gaussian fit is in dashed blue line. The red square points and solid line are from the GenFIT.
}
\label{fig:resolution_comparison}
\end{figure}

Initially, the track fitting algorithm was first implemented using true initial parameters and no backgrounds. This was done twice, first with the disc geometry and second with the straw geometry. The momentum of signals fitted with the disc geometry can be seen in Fig.~\ref{fig:resolution_comparison}. In the same figure, the results for the same measurements fitted in GENFIT2 \cite{Hoppner:2009af} are also shown. GENFIT2 is an established track reconstruction toolkit first published in 2010. It mainly focuses on track fitting, using realistic Geant4 geometry and propagation techniques. The figure shows that, in principle, ACTS is competitive with GENFIT2. 
The track fitter in ACTS achieved a momentum resolution of 236~keV/c, which is a bit bigger than the experimental requirement.

In a second run with the straw geometry, it was found that most reconstructed trajectories had less than 10 measurements, the majority having less than 5. Thus, no trajectory could be used to extract the track parameters. The shortage of measurements might occur because of the \texttt{Acts::Layer} overlap mentioned in section \ref{sec:reconstruction}. To avoid it, the \texttt{Acts::Layer}s have to be smaller than the enclosed straws, and that might affect the propagation of the fitter. At the moment, ACTS does not have an official alternative to this issue, but new geometry descriptions are being developed to solve it. Because of this, in the following section only the disc geometry is considered.

In order to represent a realistic scenario, the Kalman Fitter was implemented inputting the estimated initial parameters from seeds. The results are shown in Fig.~\ref{fig:init_mom}. The track fitter, while successfully redirecting some of the inaccurate initial estimations, is unavoidably biased. Thus, the resolution of the fit worsens, reaching values over 6~MeV/c, and many of the fitted tracks are automatically discarded. The drop in resolution is especially concerning in the presence of background hits. The resolution can be improved by setting a maximum $\chi^2/NDF$ value for the reconstructed tracks. In Fig.~\ref{fig:init_mom} the histogram considering tracks with $\chi^2/NDF<50$ is shown in red as a reference. The maximum removes most of the worse fitted tracks, and more stringent values further improve the resolution. However, in this removal many tracks with a proper set of signal electron measurements are also removed.

With background hits, the measurements belonging to the same trajectories need to be found before track fitting. This is done using the CKF. Starting from the estimated initial parameters from the seeds, the CKF finds the measurements closest to the expected trajectories. The CKF, in the absence of backgrounds, performs a simple Kalman Fitting. 

\subsection{Full reconstruction performance}
The quality of the full reconstruction algorithm is considered based on the efficiency, fake rate, and purity of the tracks. To avoid the high energy tail of DIO electrons, the allowed reconstructed momentum region for signal electrons is ($104.2<p<105.5$)~MeV/c. Thus, the efficiency of the algorithm is defined as the ratio of signal electrons found in that region with respect to the total amount of signal electrons considered. Furthermore, the reconstructed trajectories should have high purity. In this study, reconstructed events with a purity of 0.5, i.e. with 50\% of the measurements belonging to signal electrons, are considered reconstructed signals. In addition, the fake rate in the signal momentum region should be as reduced as possible. A reconstructed track is deemed ``fake" if it consists of less than 50\% of signal electron measurements.
As mentioned in Sec. \ref{sec:detector}, COMET Phase-II consists of two C-shaped solenoids that strongly suppress background particles reaching the detector, and the proton beam extinction factor of $10^{-11}$ efficiently removes the possibility of unforeseen particles. 
In consequence,
background particles are not expected in the signal momentum region, hence fake signals should be effectively removed. 

The results of the track reconstruction algorithm 
(efficiency, purity and fake rate)
are shown in Table \ref{tab:full}.
Around 10,000 signal events are considered. Each reconstruction run was performed with background particles arising from $2.7\times10^{7}$ protons hitting the pion production target.
Different values are shown based on the cuts applied to the triplets, as shown in Table \ref{tab:seeding}. The performance is shown both before and after the refitting stage. In the cases where the ACTS algorithm alone is used and together with the $t$-cut, the refitting stage is not applied due to the large amount of background measurements present in the fitted trajectories.
The results clearly show the improvement that the independent $t$- and $p$-cut make. In particular, the $p$-cut doubles the original efficiency of 5\%. Furthermore, the refitting stage also raises the efficiency from 9\% to 14\%, in the best case. The refitting stage considerably increases the purity of the final tracks, averaging at around 0.8 for all reconstructed tracks. Additionally, the fake tracks in the signal region are also further removed, dropping to $10^{-4}$.

\begin{table}[htbp]
\centering
\caption{Full reconstruction performance for different seed constraints. The performance is shown before the refitting procedure and after it when applying the $p$-cut alone and both of the constraints together. 
\label{tab:full}}
\smallskip
\begin{tabular}{p{3cm}| c |c |c c| c c}
\hline\hline
 & ACTS & ACTS + $t$-cut & \multicolumn{2}{c|}{ACTS + $p$-cut} & \multicolumn{2}{c}{ACTS + $t$-cut + $p$-cut} \\  

     & Fit & Fit & Fit & Refit & Fit & Refit \\  
\hline
Efficiency & 0.049 & 0.052 & 0.10 & 0.13 & 0.088 & 0.137 \\
Purity of all tracks & 0.43 & 0.43 & 0.56 & 0.78 & 0.58 & 0.81\\
Purity of signal $e$ & 0.65 & 0.69 & 0.79 & 0.92 & 0.81 & 0.95\\
Fake tracks & 0.003 & 0.002 & 0.001 & 0.0004 & 0.0003 & 0.0002\\
\hline\hline
\end{tabular}
\end{table}

\section{Conclusion}
\label{sec:conclusion}
The COMET experiment will search for muon to electron conversion, a charged lepton flavour violating process. The signal of this conversion process is a mono-energetic electron of 104.97~MeV/c. To search for it, the Phase-II of COMET will use a combination of a straw tracker and an electron calorimeter detector. The straw tracker will be used to reconstruct the trajectories of the particles through the detector. An efficient and accurate track reconstruction algorithm is required to discern the signal electrons from the background. In this study, we have implemented the ACTS tracking software into the COMET Phase-II straw tracker. This represents the first time that ACTS has been used for multi-turn fitting in longitudinal detectors and in the 100~MeV/c momentum region. 

The straw tracker was modelled in ACTS as a simplified geometry containing discs instead of straw layers, and as a detailed geometry with all of the straws. The latter geometry yielded a suboptimal fitting performance. This might be due to the current inability of embedding staggered straw layers within the \texttt{ACTS:Layer}s. Work is being done within the ACTS collaboration to solve this issue. However, in this study, we focused on the performance using the simplified disc geometry since it gave more reliable results. With the disc geometry, the signal electron fitting had a resolution of around 200~keV/c when the initial parameters were the true ones. This was an encouraging result, especially considering the usage of a simplified geometry.

The fitting algorithm was tested by using the initial parameter estimator included in ACTS. This estimator uses triplets of measurements to estimate the parameters of the trajectory. The estimator showed an important underestimation of the momentum, which was reflected in the fitting results. The Kalman Fitter was able to improve the estimation, but the deviation of the fitted momentum with respect to the true one was too large for most cases. A $\chi^2/NDF$ cut can be applied to improve it, but these results highlight the need for a better parameter estimator for trajectories in the 100~MeV/c momentum region.

The track seeding algorithm in ACTS was used alongside two dedicated constraints, $t$-cut and $p$-cut, to remove background hits. The seeding algorithm had an efficiency of 82\%, which dropped to 74\% when using the two constraints. However, the additional constraints removed up to 62\% of the fake and background seeds. This resulted in an improved full reconstruction efficiency. The full reconstruction efficiency raised from 5\% to 10\% by using the $p$-cut, and the purity of the reconstructed signal electrons reached 0.81 when using both cuts together. In the final refitting stage, the efficiency reached 14\% and the purity of reconstructed signals was 0.95 when using both dedicated constraints together. In that case, 0.02\% fake events were found. These fake events can be recognised by their large $\chi^2$ sum value, but their removal is left for a future study. 

The achieved efficiency using ACTS is too low for the goals of COMET Phase-II, which aims at an efficiency close to 80\%. The first cause for the low efficiency is the inability to use an accurate geometry description in ACTS. In addition, the initial parameter estimator is not accurate enough to initialise tracks in the 100~MeV/c region. In this study, we have shown a first approach to implementing ACTS in lower energy regions for multi-turn tracking. In the future, more background events and combined usage of the calorimeter with the straw tracker should be considered.

\acknowledgments
This work was supported by the National Research Foundation of Korea (NRF) grant No. 2022R1F1A1060075 funded by the Korea government (MSIT), and the BK21 FOUR funded by the Ministry of Education (MOE, Korea) and National Research Foundation of Korea (NRF).
We thank KEK and J-PARC, Japan for their support of infrastructure and the operation of COMET. Crucial computing support from all partners is gratefully acknowledged, in particular from CC-IN2P3, France; GridPP, United Kingdom; and Yandex Data Factory, Russia, which also contributed expertise on Machine Learning methods. 
The authors acknowledge the help and discussions of the ACTS team, especially from Dr. B.K.Yeo (U.C. Berkeley).

\bibliographystyle{JHEP}
\bibliography{biblio.bib}

\end{document}